\documentclass[epj]{svjour}
\usepackage{latexsym,graphicx,epsfig,psfrag,here,color,citesort}    
\def\lesssim{\mathrel{\hbox{\rlap{\hbox{\lower4pt\hbox{$\sim$}}}\hbox{$<$}}}}
\def\query#1{\marginpar{\begin{flushleft}\footnotesize#1\end{flushleft}}}

\begin{document}

\def\eq{\begin{eqnarray}}
\def\en{\end{eqnarray}}

\def\query#1{\marginpar{\begin{flushleft}\footnotesize#1\end{flushleft}}}%

\title{$\bar KN$ scattering lengths from the experiments on kaonic hydrogen
and kaonic 
deuterium}
\author{Akaki Rusetsky\inst{1,2}}
\institute{
Universit\"{a}t Bonn, Helmholtz-Institut f\"{u}r Strahlen- und 
Kernphysik (Theorie),\\ 
Nu\ss allee 14-16, D-53115 Bonn, Germany
\and
On leave of absence from: High Energy Physics Institute,
Tbilisi State University,\\
University St.~9, 380086 Tbilisi, Georgia
}
\date{Received: date / Revised version: date}
%
\abstract{
Within the framework of a low-energy effective field theory 
we consider the procedure of extraction of the S-wave kaon--nucleon scattering 
lengths $a_0$ and 
$a_1$ from a  combined fit to the kaonic hydrogen and 
kaonic deuterium data. 
It is demonstrated
 that, if the present DEAR central values for the kaonic
hydrogen ground-state energy and width are used in the
analysis of the data, a solution for $a_0$ and $a_1$ exists 
only in a restricted domain of input values for the kaon-deuteron scattering 
length. We therefore conclude that forthcoming measurement of 
this scattering length imposes
stringent constraints on the theoretical description of the kaon-deuteron
interactions at low energies. Most of the results of this talk are contained
in the recent paper~\cite{Raha4}.
}

\maketitle

\section{Introduction}
\label{sec:intro}

Recently, the DEAR collaboration at LNF-INFN has performed a measurement
of the energy level shift and width of the kaonic hydrogen ground 
state~\cite{Beer} with a considerably better accuracy than the earlier
KpX experiment at KEK~\cite{KEK}. The preliminary result of DEAR is
\eq
\epsilon_{1s}&=&193\pm 37~\mbox{(stat)}\pm 6~\mbox{(syst) eV}\, ,
\nonumber\\[2mm]
\Gamma_{1s}&=&249\pm 111~\mbox{(stat)}\pm 30~\mbox{(syst) eV}\, .
\en
Now DEAR is being replaced
by the SIDDHARTA experiment, which by 2007 plans measurements
of both the energy shift and the
width of kaonic hydrogen with a precision of several
eV, i.e. at the few percent level. Moreover, 
SIDDHARTA will attempt the
first ever measurement of the energy shift of the kaonic deuterium 
with a comparable accuracy and
possibly, of other atomic complexes (kaonic helium and sigmonic atoms).

The necessity to perform  measurements of the kaonic deuterium ground-state
observables is evident from the following simple argument. Namely,
the measurement of only the kaonic hydrogen spectrum does not allow one
-- even in principle -- 
to extract independently both S-wave $\bar KN$ scattering lengths $a_0$ and 
$a_1$. The reason for this is that
there are open inelastic channels below the $\bar KN$ threshold,
rendering these scattering lengths complex. Consequently, one has to
determine four independent quantities (real and imaginary parts of $a_0$
and $a_1$) that requires performing four independent measurements -- e.g., the 
energy level shifts and widths of kaonic hydrogen {\em and} 
kaonic deuterium. However, even though it is clear that $a_0$ and $a_1$ can 
not be determined separately without measuring kaonic deuteron, it is still 
not {\it a priori} evident, whether it is possible to do so if one performs 
such a measurement. Note that to this end one needs theoretical input
from three-body calculations, which relate $a_0$ and $a_1$ to the observables
of the kaonic deuterium and one has to check,
whether the uncertainties in these calculations are small 
enough not to hinder a determination of $a_0$ and $a_1$
from the outcome of the experiment.

A time-honored phenomenological 
approach to the ka\-on-deuteron interactions at low energies is
based of Faddeev equations in the potential scattering theory,
see, e.g.
\cite{Hetherington,Toker,Torres:1986mr,Deloff:1999gc,Barrett:1999cw,Bahaoui}.
For the recent status of the problem, see e.g. \cite{Gal-conf} and references therein. In this
approach, one predicts the numerical value of the $K^-d$ scattering length,
using ``realistic'' input two-body potentials in the calculations. On the
other hand,  one may derive an expression for the
$K^-d$ scattering length in a form of the (partially re-summed)
multiple-scattering series, which
algebraically relates this scattering length to $a_0$ and $a_1$ and which 
thus can be used to solve the problem of extracting $a_0$ and $a_1$ 
from the experiment (inverse problem). 
We stress that within the potential picture
 one may always refer to the exact
solution of Faddeev equations for a numerical check of the validity of the 
multiple-scattering expansion which, in most cases, works reasonably well
\cite{Gal-conf}.

In recent years the investigations of the same problem within the effective
field theory (EFT) framework have 
started to appear (see e.g.~\cite{Kamalov:2000iy,Raha4}). 
An ultimate
 goal of these investigations is to perform a model-independent
calculation of the
kaon-deuteron scattering observables without referring to the exact form
of phenomenological hadronic interactions.  
Moreover, this type of the approach has the potential to go
 beyond the approximations that
have been used to derive the multiple-scattering series within the potential
model. Namely, one may expect that using the EFT methods could
allow one 
to systematically improve the accuracy of the calculations by including
e.g. the effects, coming
from higher-order terms in the effective-range expansion
of the $\bar KN$ amplitudes. Other effects like the nucleon recoil, or the
short-range three-body forces, should also be taken into account, if one aims
at a high accuracy in final results, which should be compared to the precise
data coming from future experiments.

In our opinion, at present stage of the studies
 potential models provide useful testing ground for the validity
of various schemes in EFT, since the relative size of different
contributions can be directly estimated there by using numerical methods.
In particular, the non-relativistic EFT, which we shall be using below,
has a structure very similar to the potential model and leads to the same
multiple-scattering series at leading order. 
On the other hand, a straightforward generalization
of this approach leads to the inclusion of the
 higher-derivative local interactions in
the kaon-nucleon Lagrangian, corresponding to the ef\-fec\-ti\-ve-range expansion
of the $\bar K N$ amplitudes, as well as of the various terms describing relativistic
corrections and short-range three-body interactions. 
It will be challenging to carry out a
systematic quantitative analysis of these (sub-leading) effects,
including the analysis of the theoretical error, with an aim to observe
a consistent improvement of the theoretical precision. Note that, 
in order to be able to compare with the future accurate data coming from
SIDDHARTA experiment, such calculations should be necessarily performed.

The main difference of the present study of the kaonic deuterium
 to previous work consists in the following.
The existing approaches were exclusively concentrated on the prediction
of the $K^-d$ scattering length from the input $\bar KN$ scattering lengths.
We are not aware of the ``reversed'' analysis in the literature, where the
$\bar KN$ scattering lengths are determined from the input data of kaonic
hydrogen and deuterium ground-state shift and width.
However, this is exactly the type of the analysis that
will be required in the near future 
for the SIDDHARTA data. In the present work we provide such an analysis.
Moreover, in the
absence of any experimental input for the deuteron, 
we argue in favor of using ``synthetic''
 data instead: just scanning the complex plane 
($\mbox{Re}\,a_{Kd},\,\mbox{Im}\,a_{Kd}$) in a reasonable interval and
using these values of the kaon-deuteron scattering length together with the
 $K^-p$ elastic scattering length, which is measured in the kaonic hydrogen
 experiment, for extracting the values of $a_0$ and $a_1$.
In this way we demonstrate that
the reversed calculations, owing to the non-linear 
dependence of the kaon-deuteron amplitude on the $\bar KN$ scattering 
lengths, turn out to be much more sensitive to the theoretical input on
the deuteron structure and the kaon-deuteron interactions, 
than a straightforward evaluation of the $K^-d$ scattering
length through the multiple-scattering series. This fact could potentially
render a combined analysis of the hydrogen and deuterium data a beautiful
testing ground for different EFT descriptions of the 
low-energy kaon-deuteron interactions and, 
as a result, might enable one to accurately
determine the values of the scattering lengths $a_0$ and $a_1$.

\section{Kaonic hydrogen and kaonic deuterium}
\label{sec:deuterium}

In the experiments on hadronic atoms one measures the energy levels and
widths of this sort of bound states. At present, there exists a well 
established systematic procedure for extracting 
the values of the pertinent hadronic 
scattering amplitudes at threshold from these measurements, based on
non-relativistic effective Lagrangians (see e.g. 
\cite{pipi,Bern,piK,piN,Raha1,Raha2,Raha4}). Below we merely present the result
for the kaonic hydrogen and kaonic deuterium without a derivation.
The details can be found in Refs.~\cite{Raha1,Raha2,Raha4}.

From the measurement of the (complex) energy shift of the hadronic atoms
the elastic threshold amplitudes can be extracted by using following 
relations
\eq\label{eq:Deser_kp}
\epsilon_{1s}&-&i\,\frac{\Gamma_{1s}}{2}=-2\alpha^3\mu_c^2\,
a_p
\\[2mm]
&\times&\bigl\{1-2\alpha\mu_c\,a_p\,(\ln\alpha-1)
+\cdots\bigr\}\, ,
\nonumber\\[2mm]
\label{eq:Deser_kd}
\epsilon_{1s}^d&-&i\,\frac{\Gamma_{1s}^d}{2}=-2\alpha^3\mu_r^2\,
A_{Kd}
\nonumber\\[2mm]
&\times&\bigl\{1-2\alpha\mu_r\,A_{Kd}\,(\ln\alpha-1)
+\cdots\bigr\}\, .
\en
Here, $\epsilon_{1s},\,\Gamma_{1s}$ and $\epsilon_{1s}^d,\,\Gamma_{1s}^d$
are the ground-state strong shift and the width of the kaonic hydrogen and
 kaonic deuterium,
respectively. Further, $a_p$ and $A_{Kd}$ denote threshold amplitudes
for the processes $K^-p\to K^-p$ and $K^-d\to K^-d$, and $\mu_c$, $\mu_r$
denote the reduced masses of the $K^-p$ and $K^-d$ bound systems, respectively.
In the following, we shall also refer to $A_{Kd}$  as to the ``kaon-deuteron
scattering length.''

The above universal relations are exact at next-to-leading order in the
isospin-breaking parameters: the fine-structure constant $\alpha$ and the 
up- and down-quark mass difference $m_d-m_u$. The threshold amplitudes $a_p$ 
and $A_{Kd}$ contain, by definition, isospin-breaking corrections up to
and including $O(\alpha,(m_d-m_u))$.

The analysis of the data proceeds as follows.
In the experiment, one measures two complex quantities $a_p$ and $A_{Kd}$.
At the next step, one expresses $a_p$ and 
$A_{Kd}$ in terms of $a_0$ and $a_1$ in order to extract their values from the
experiment. In the case of $A_{Kd}$ this implies addressing
three-body problem in the kaon-two-nucleon sector (see below).
Note also that from now on we omit virtual photons: they were
primarily needed to create bound states of kaons with the proton and the 
deuteron. We believe that at the accuracy of presently available
calculations, virtual photon corrections to hadronic observables can be safely
neglected.

\section{Multiple-scattering series for the kaon-deuteron amplitude}
\label{sec:multiple}

The effective field theory, which is used to describe $K^-d$ scattering,
is constructed in the following manner:
\begin{itemize}
\item[a)]
The $\bar KN$ sector is described by the non-relativistic effective Lagrangian
of the type considered in Ref.~\cite{Raha1} (without virtual photons)
\eq\label{Lagr-ini}
{\mathcal L}_{\bar KN}&=&\psi^\dagger\,\biggl\{i\partial_t-m_p+\frac{\nabla^2}{2m_p}
+\cdots\biggr\}\,\psi
\nonumber\\[2mm]
&+&\chi^\dagger\,\biggl\{i\partial_t-m_n+\frac{\nabla^2}{2m_n}
+\cdots\biggr\}\,\chi
\nonumber\\[2mm]
&+&\sum_\pm (K^\pm)^\dagger\,\biggl\{i\partial_t-M_{K^+}
+\frac{\nabla^2}{2M_{K^+}}+\cdots\biggr\}\,K^\pm
\nonumber\\[2mm]
&+&{(\bar K^0)}^\dagger\,\biggl\{i\partial_t
-M_{\bar  K^0}+\frac{\nabla^2}{2M_{\bar K^0}}
+\cdots\biggr\}\,\bar K^0
\nonumber\\[2mm]
&+&\tilde d_1\,\psi^\dagger\psi\,(K^-)^\dagger K^-+
\tilde d_2(\psi^\dagger\chi\,(K^-)^\dagger\bar K^0+h.c.)
\nonumber\\[2mm]
&+&\tilde d_3\chi^\dagger\chi\,(\bar K^0)^\dagger\bar K^0
+\tilde d_4\chi^\dagger\chi\,(K^-)^\dagger K^-+\cdots\, ,
\en
where $\psi$, $\chi$, $K^\pm$ and $\bar K^0$ stand for non-relativistic
proton, neutron, $K^\pm$ and $\bar K^0$ fields, respectively,
$m_p$, $m_n$, $M_{K^+}$ and $M_{\bar K^0}$ denote the masses of these particles
and $\tilde d_i,~i=1,2,3,4$ are expressed through the threshold 
scattering amplitudes in pertinent channels, reducing to certain combinations
of $a_0$ and $a_1$ in the isospin limit. The inclusion of derivative 
interactions, corresponding to the higher-order terms in the effective-range
expansion of the $\bar KN$ interactions, is straightforward.

\item[b)] The interactions between the nucleons are described
in ChPT with non-perturbative pions (for a recent review, see 
e.g.~\cite{Epelbaum-review}). In the effective theory, which is used here,
the $\bar KN$ and $NN$ sectors do not talk to each other, by construction.
Therefore, the only input that one needs from the $NN$ sector is that
the nucleon-nucleon
potential in the momentum space in CM frame is given by a known function 
$V_{NN}({\bf p},{\bf q};\Lambda)$, where $\Lambda$ denotes the cutoff parameter
in this scheme -- typically, of order of a few hundred MeV (to ease notations, 
we suppress spin-isospin indices). This potential leads to the formation of
a shallow bound state in the $^3S_1$ channel (deuteron), with the wave function
$\Psi_d({\bf p};\Lambda)$.

\item[c)] Three-body interactions in the kaon-two-nucleon sector are again
described by a local Lagrangian of the type
\eq\label{eq:L3}
{\cal L}_{\bar KNN}=\tilde f_0\,\psi^\dagger\psi\,\chi^\dagger\chi\, 
(K^-)^\dagger K^-+\cdots\, ,
\en
where ellipses stand for the terms with derivatives.

\item[d)] The kaon-deuteron scattering amplitude can be calculated in a 
standard manner, evaluating the six-point function $\bar KNN\to\bar KNN$
and extracting the double pole, corresponding to the initial and 
final deuterons (see, e.g.~\cite{Raha2}). 
The multiple scattering series similar to that of Ref.~\cite{Kamalov:2000iy}
are obtained under usual approximations:
\begin{itemize}
\item
Diagrams, containing
 iterations of $V_{NN}$ within the loops, are omitted.
\item
The Fixed Center Approximation (FCA) is used to simplify propagators in the
Feynman diagrams: the kinetic energies of the nucleons are neglected.
\item
In addition, derivative interactions in the $\bar KN$ and $\bar KNN$ sectors 
are neglected.
\item
Relativistic corrections for kaons are not included.
\end{itemize}
Note that the validity of FCA
been studied both in the potential scattering theory 
(see e.g.~\cite{Faldt,Bahaoui})
and in the EFT approach~\cite{Hanhart}. 
In Ref.~\cite{Kamalov:2000iy} (see also references therein) it is argued
that FCA can be a reasonable
approximation even for $M_{K^+}/m_p\simeq 0.5$.
This fact should be related to the peculiar cancellations 
at second order, which are discussed in Refs.~\cite{Faldt,Hanhart}.
We could also observe a clear pattern of such cancellations in our 
exploratory 
calculations. Further, even if there exists no proof of cancellations beyond
second order, from e.g. the comparison to the exact solution of Faddeev
equations~\cite{Deloff:1999gc} 
(see also the discussion in Ref.~\cite{Kamalov:2000iy})
one may conclude that FCA works reasonably well also for the re-summed
multiple-scattering series (Note, however Ref.~\cite{Bahaoui}, 
where it has been pointed out that
large corrections to FCA might emerge due to the presence of the 
nearby sub-threshold resonance in the $\bar KN$ channel.).

\item[e)]
\begin{sloppypar}
We wish to note that in the case of the pion-deuteron scattering, the first few terms 
of the above
multiple-scattering expansion can also be derived by using ChPT with 
non-perturbative pions up to and including $O(p^4)$~\cite{Bernard}. Additional
small terms (e.g. boost correction) correspond to the derivative contribution
in the meson-nucleon Lagrangian. In the case of the $\bar KN$ scattering,
such a comparison to ChPT is not possible for obvious reasons.
\end{sloppypar}

\item[f)]
As already discussed in Ref.~\cite{Gal-conf}, a crucial difference between the
pion-deuteron and kaon-deuteron cases is that the multiple scattering series 
diverges for the latter whereas it converges for the former (the $\bar KN$
scattering lengths are large). For this reason, one has to perform a (partial)
re-summation of the multiple-scattering series. Under the approximations
listed below, this can be easily done.

\end{itemize}

Finally, we arrive at the following expression
for the pion-deuteron scattering length
\eq\label{eq:final-Kamalov}
\biggl(1+\frac{M_{K^+}}{M_d}\biggr)A_{Kd} =\int_0^\infty dr\,(u^2(r)+w^2(r))\,
\hat a_{kd}(r)\, ,
\en
where $M_d$ is the deuteron mass, $u(r)$ and $w(r)$ denote the usual 
$S-$ and $D-$wave components of the deuteron
wave function $\Psi_d({\bf r};\Lambda)$ and the normalization condition 
is given by
$\int_0^\infty dr\,(u^2(r)+w^2(r))=1$. Further,
\eq\label{eq:ratio-Kamalov}
\hat a_{kd}(r)=\frac{\tilde a_p+\tilde a_n
+(2\tilde a_p\tilde a_n-b_x^2)/r-2b_x^2\tilde a_n/r^2}
{1-\tilde a_p\tilde a_n/r^2+b_x^2\tilde a_n/r^3}+\delta \hat a_{kd}
\nonumber\\
\en
with $b_x^2=\tilde a_x^2/(1+\tilde a_u/r)$. Furthermore,
\eq
\biggl(1+\frac{M_{K^+}}{m_p}\biggr)a_{p,n,x,u}=\tilde a_{p,n,x,u}\, ,
\en
where $a_{p,n,x,u}$ denote the threshold scattering ampli\-tu\-des
for $K^-p\to K^-p$,  $K^-n\to K^-n$, $K^-p\to \bar K^0n$ and 
$\bar K^0n\to \bar K^0n$, respectively (these quantities are proportional
to pertinent $\tilde d_i,~i=1\cdots 4$ from Eq.~(\ref{Lagr-ini})). Finally, the quantity
$\delta \hat a_{kd}$ is proportional to the three-body LEC
$\tilde f_0$ from Eq.~(\ref{eq:L3}). The value of this constant is
completely unknown at present. Different estimates 
(e.g. the dimensional estimate
or the study of the $\Lambda$-dependence) lead to the conclusion that the 
uncertainty introduced by this term is not very large. For this reason,
we further neglect this contribution altogether. In this case, 
Eqs.~(\ref{eq:final-Kamalov}) and (\ref{eq:ratio-Kamalov}) become formally 
identical to ones derived in Ref.~\cite{Kamalov:2000iy}.

\section{Isospin breaking}
\label{sec:isospin}

The equation~(\ref{eq:ratio-Kamalov}) contains four different 
combinations of the threshold amplitudes. Consequently, one has first
to express these amplitudes in terms of two scattering lengths $a_0$ and
$a_1$, which should then be determined from the analysis of the
combined data on  kaonic hydrogen and deuterium. Avoiding this step and
working in the particle basis is not possible: in this case, number of unknowns
$a_p,a_n,a_x,a_u$ exceeds the number of independently measured observables.

We take into account the leading-order isospin-bre\-a\-king
corrections in the kaon-nucleon scattering 
amplitudes which are due to the unitary cusps~\cite{Raha1}.
The re-summation of the bubble diagrams leads to the following simple 
parameterization 
\eq\label{eq:cusps}
a_p&=&\frac{\frac{1}{2}\,(a_0+a_1)+q_0a_0a_1}{1+\frac{q_0}{2}\,(a_0+a_1)}\, ,
\quad
a_n\!=\!a_1\, ,
\nonumber\\
a_x&=&\frac{\frac{1}{2}\,(a_0-a_1)}{1-\frac{iq_c}{2}\,(a_0+a_1)}\, ,
\quad
a_u\!=\!\frac{\frac{1}{2}\,(a_0+a_1)-iq_ca_0a_1}{1-\frac{iq_c}{2}\,(a_0+a_1)}\, ,
\nonumber\\
&&
\en
where
\eq
&&q_c=\sqrt{2\mu_c\Delta}\, ,\quad
q_0=\sqrt{2\mu_0\Delta}\, ,
\nonumber\\[2mm]
&&\Delta=m_n+M_{\bar K^0}-m_p-M_{K^+}\, ,
\nonumber\\[2mm]
&&\mu_c=\frac{m_pM_{K^+}}{m_p+M_{K^+}}\, ,\quad
\mu_0=\frac{m_nM_{\bar K^0}}{m_n+M_{\bar K^0}}\, .
\en
Bearing in mind large isospin-breaking corrections to the individual amplitudes
in Eq.~(\ref{eq:cusps}), one may wonder whether the isospin-breaking correction
to the kaon-deuteron scattering length, which is determined from 
Eqs.~(\ref{eq:final-Kamalov}) and (\ref{eq:ratio-Kamalov}), is also large.
Our present calculations have, however, shown that this is not the case.

\section{Numerical results and discussion}
\label{sec:numerics}

\begin{figure}[t]
\begin{center}
\vspace*{.7cm}
\includegraphics[width=8.cm]{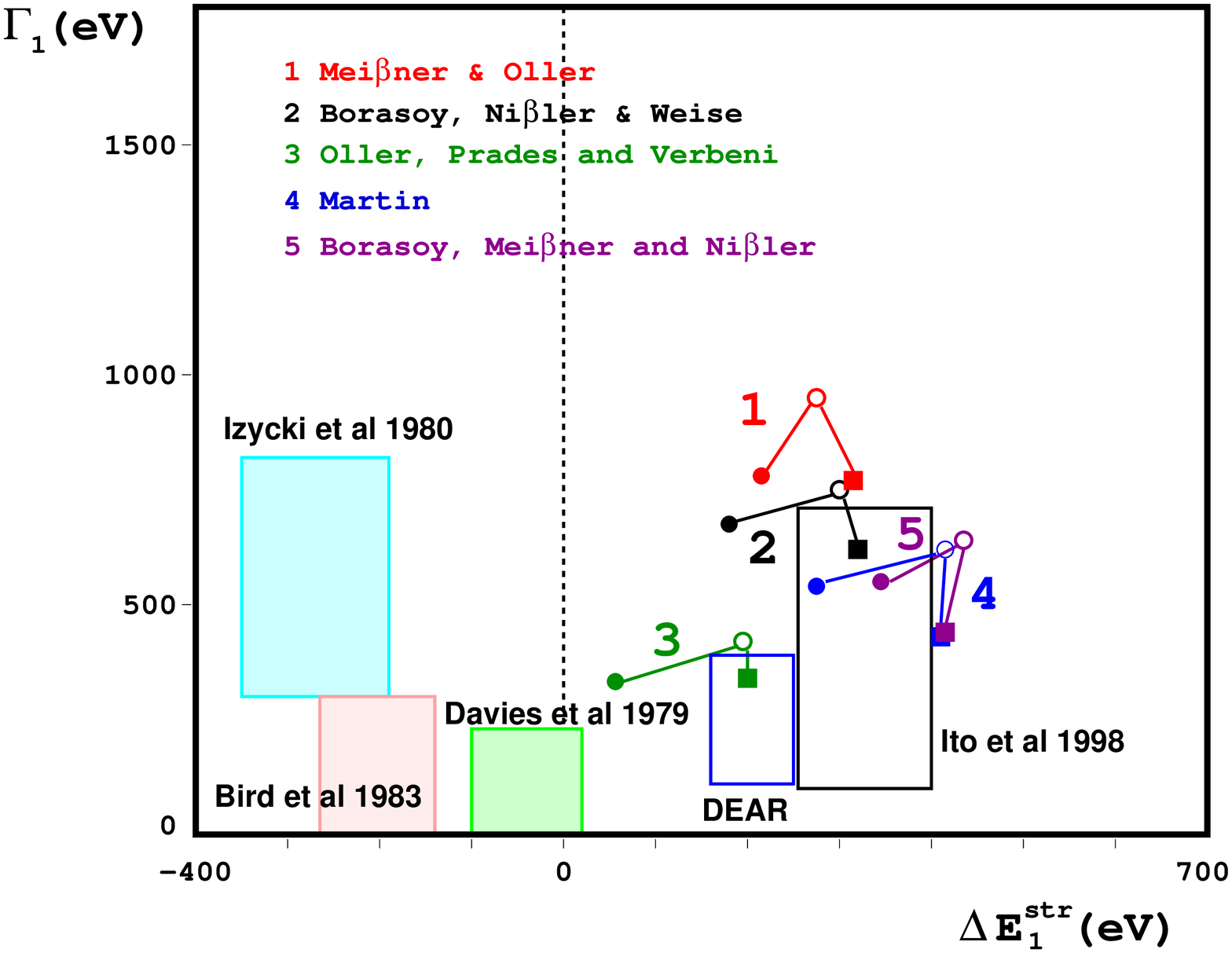}
\end{center}
\caption{Comparison of the calculated values of the strong shift and width in
kaonic hydrogen~\cite{MO,BNW,OPV,Martin,BMN} 
with existing experimental data. Filled circles correspond to completely 
neglecting isospin-breaking corrections. Empty circles correspond to
including isospin breaking in $a_p$, according to Eq.~(\ref{eq:cusps}) but
neglecting the Coulomb corrections in Eq.~(\ref{eq:Deser_kp}) (the term, proportional to $\alpha(\ln\alpha-1)$ in curly brackets). Finally, filled squares correspond to the full result in Eq.~(\ref{eq:Deser_kp}).}
\label{fig:square}
\end{figure}

The comparison of the results for kaonic hydrogen to experiment is 
displayed in Fig.~\ref{fig:square}. Here, we have used the values of $a_0$ and
$a_1$ given in Refs.~\cite{MO,BNW,OPV,Martin,BMN} as theoretical input
in Eq.~(\ref{eq:Deser_kp}). On the basis of this comparison, one may conclude
that the scattering lengths that are obtained from the fit to the data above
threshold, in most of the cases are not compatible with the DEAR measurement.
It is also seen that the isospin-breaking corrections are huge and can
not be neglected even at the present accuracy of the experiment.

\begin{figure}[t]
\begin{center}
\vspace*{.7cm}
\includegraphics[width=8.cm]{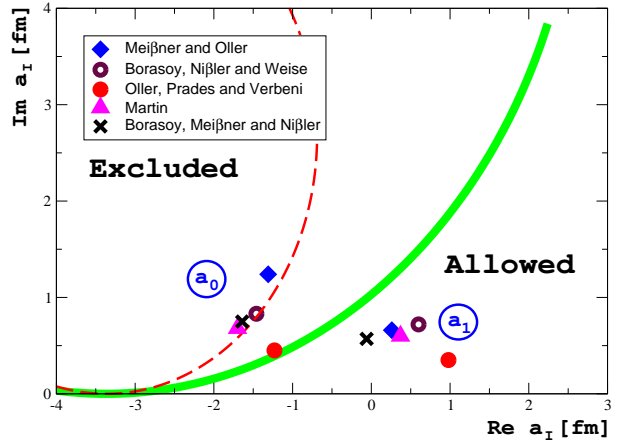}
\end{center}
\caption{Restrictions set by the DEAR data on the values of the scattering 
lengths $a_0$ and $a_1$ (thick solid line). For comparison, we give the scattering length 
calculations from different analyzes: 
1) Mei{\ss}ner and Oller~\cite{MO}, 
2) Borasoy, Ni{\ss}ler and Weise, fit u~\cite{BNW},
3) Oller, Prades and Verbeni, fit A4~\cite{OPV},
4) Martin~\cite{Martin},
5) Borasoy, Mei{\ss}ner and Ni{\ss}ler~\cite{BMN}. 
The dashed line corresponds to the restrictions, obtained by using 
KpX data instead of DEAR data.}
\label{fig:circle}
\end{figure}

The compatibility of different values of $a_0$ and $a_1$ with the experiment
can be made more transparent with the help of the following picture.
From Eq.~(\ref{eq:cusps}) it is seen that, for a given $a_p$ the (complex)
quantities $a_0$ and $a_1$ obey the equation
\eq\label{eq:circle}
a_0+a_1+\frac{2q_0}{1-q_0a_p}\,\, a_0a_1-\frac{2a_p}{1-q_0a_p}=0\, .
\en
Together with the requirement ${\rm Im} \,a_I\geq 0$, which stems from
unitarity, Eq.~(\ref{eq:circle}) defines a circle in the
(${\rm Re}\,{a_I}$, ${\rm Im}\,{a_I}$)--plane. Part of this circle is shown
in Fig.~\ref{fig:circle} (note that, bearing in mind the preliminary 
character of the DEAR data \cite{Beer}, we use only central values
in order to illustrate the construction of
the plot and do not provide a full error analysis). 
In order to be consistent with the DEAR data, both $a_0$ and $a_1$ should be
on the right of this universal DEAR circle. For comparison, on the same figure
we also indicate the (much milder) restrictions, which arise, when the KpX data
are used instead of DEAR data. As we see,
in most of the approaches it is rather
problematic to get a value for $a_0$ which is compatible with DEAR.
This kind of analysis may prove useful in the near future,
when the accuracy of
the DEAR is increased that might stir  efforts on the theoretical side,
aimed at a systematic quantitative description of the $\bar KN$ 
interactions within the unitarized ChPT.

\begin{figure}[t]
\begin{center}
\vspace*{.7cm}
\includegraphics[width=8.cm]{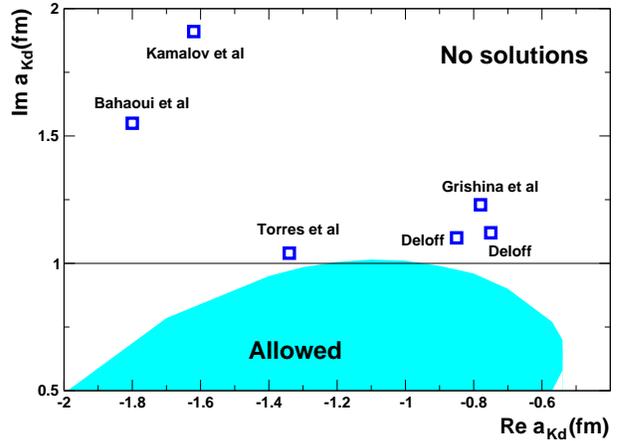}
\end{center}
\caption{The region in the $({\rm Re}\,A_{Kd}$,  ${\rm Im}\,A_{Kd})$--plane
where  solutions for $a_0$ and $a_1$ exists.
For comparison, we also show the results of
calculations of the kaon-deuteron scattering length from 
Refs.~\cite{Kamalov:2000iy,Torres:1986mr,Deloff:1999gc,Bahaoui,Grishina}.}
\label{fig:area}
\end{figure}

\begin{sloppypar}
Finally, we turn to our main goal of determining both $a_0$ and $a_1$
from the simultaneous analysis of the kaonic hydrogen and kaonic deuterium 
data. In the beginning, 
from Eq.~(\ref{eq:circle}) one may determine e.g. $a_1$.
Substituting this expression into Eqs.~(\ref{eq:final-Kamalov}),
(\ref{eq:ratio-Kamalov}) and (\ref{eq:cusps}), one arrives at a 
non-linear equation\footnote{In Eq.~(\ref{eq:final-Kamalov}) we use the 
NLO wave function of the deuteron with the cutoff parameter 
$\Lambda=600~\mbox{MeV}$~\cite{Epelbaum-private}. The final results however show a very weak dependence
on the cutoff.} for determining $a_0$ with a given input value
of $A_{Kd}$.
In the absence of  experimental data on kaonic deuterium, 
we have scanned the $({\rm Re}\,A_{Kd}$,  
${\rm Im}\,A_{Kd})$--plane in the interval $-2~{\rm fm}< {\rm Re}\,A_{Kd}<0$
and  $0.5~{\rm fm}< {\rm Im}\,A_{Kd}<2.5~{\rm fm}$ 
and tried to find
solutions, using in addition DEAR input data. The results of this investigation, which are
displayed in Fig.~\ref{fig:area}, are very
interesting: it turns out that the solutions exist only 
if ${\rm Im}\,A_{Kd}\lesssim 1~{\rm fm}$ and moreover, if 
${\rm Im}\,A_{Kd}\simeq 1~{\rm fm}$ then one finds solutions only
 in a very small interval around ${\rm Re}\,A_{Kd}\simeq -1~{\rm fm}$. 
We wish to also note that all
this agrees with the scattering data analysis, carried out in 
Ref.~\cite{Sibirtsev}.
 Note that if ${\rm Im}\,A_{Kd}$ crosses the border of the shaded area in 
Fig.~\ref{fig:area} continuously from below, then on the same branch 
one gets the solution with ${\rm Im}\,a_1\leq 0$ that is forbidden 
by unitarity.
\end{sloppypar}

In conclusion we note that the region, where the solutions exist,
is much larger in the case of the KpX input than for the DEAR
input for the kaonic hydrogen. Namely, in the case of the 
KpX input the shaded area in Fig.~\ref{fig:area} covers the most part of
the plot.

\section{Conclusions}
\label{sec:concl}

Up to now, in the theoretical description of the deuteron we have restricted
ourselves to the approximations outlined above: FCA, no derivative 
interactions, no relativistic corrections, no iterations of $V_{NN}$ in the loops.
From the comparison with the potential model calculations one may expect
that these
approximations provide a good starting point for the description of the
kaon-deuteron scattering length.

\begin{sloppypar}
Within these approximations it turns out that the 
com\-bi\-ned analysis of DEAR/SIDDHARTA
data on kaonic hydrogen and deuterium 
is more restrictive than one would {\it a priori} expect. In particular,
we see that  solutions exist
only in a rather small area of the $({\rm Re}\,A_{Kd}$, 
${\rm Im}\,A_{Kd})$--pla\-ne.
Due to this fact, in certain cases it might be possible to pin down 
 the values of $a_0$ and $a_1$ at a 
reasonable accuracy, even if $A_{Kd}$ itself is not measured very accurately. 
This statement constitutes our main result.
\end{sloppypar}

Finally, we wish to emphasize that,
in the view of the forthcoming SIDDHARTA experiment, the question of 
the corrections to the leading-order approximate result acquires 
crucial importance -- even if, as expected,
they are moderate and do not change the qualitative picture.
In our opinion, it will be very useful to carry out a systematic 
calculation of these
corrections in the effective field theory framework, e.g. in the one
 described in the present 
work.

\begin{sloppypar}
{\it Acknowledgments:}
The author wishes to thank the organizers of the HYP2006 conference for 
warm hospitality and the
great effort that created an atmosphere of intense discussions and learning.
Partial financial support from the EU Integrated Infrastructure
Initiative Hadron Physics Project (contract number RII3-CT-2004-506078)
and DFG (SFB/TR 16, ``Subnuclear Structure of Matter'') is gratefully
acknowledged.
This work was supported in part by the EU Contract No. MRTN-CT-2006-035482,
\lq\lq FLAVIAnet''.

\end{sloppypar}

\end{document}